# Optical air waveguides in strong turbulence

A.GOFFIN[1,2], G.I. BABIĆ[2], AND H.M. MILCHBERG[2*]

[1]Physical Chemistry and Applied Spectroscopy, Los Alamos National Laboratory, Los Alamos, New Mexico 87545, USA
[2]Institute for Research in Electronics and Applied Physics, University of Maryland, College Park, Maryland 20742, USA
*milch@umd.edu

**Abstract:** Filament-generated air waveguides offer a route to low loss long-distance optical transport, but their viability in atmospheric turbulence has remained uncertain. Here, we demonstrate with experiments and simulations that guided beams in air waveguides can overcome prior estimated turbulence limits by orders of magnitude in the refractive index structure parameter $C_n^2$; turbulent index fluctuations much larger than the waveguide index contrast average out along the propagation path. Instead, turbulence imposes two distinct limits: scattering of the guided field beyond the waveguide acceptance aperture and, more importantly at atmospheric strengths, scintillation of the waveguide-forming beam. Scaling of experimentally validated models indicates that filament-generated air waveguides remain viable over kilometer-scale paths under strong near-ground turbulence. Our results establish turbulence limits for long-range optical guiding in the atmosphere.

## 1. Introduction

Laser filamentation occurs when a short, high-peak-power laser pulse undergoes nonlinear self-focusing and self-guided propagation in a medium, generating long-distance high-intensity, narrow channels of light and plasma [1,2]. In air, the pulse self-focuses owing to the nonlinear response of air molecules: the near-instantaneous response of the electronic orbitals and alignment of the molecular axis in the strong laser electric field. For pulses with peak powers above a critical power [3,4] $P_{cr} \sim 3.77\lambda^2/(8\pi n_{2,eff} n_0)$ in air, where $n_{2,eff}$ is the pulsewidth-dependent nonlinear refractive index [5,6], self-focusing overcomes linear diffraction and the beam collapses until arrested by field-induced plasma generation, which defocuses the pulse. The dynamic interplay between self-focusing and plasma defocusing leads to a high-intensity filament core, of diameter $d_{core} \sim 200$ μm and peak intensity $I \sim 50 - 100$ TW/cm$^2$, which exchanges energy with a lower intensity "reservoir" region [7] surrounding it, with energy flow mediated by toroidal spatiotemporal optical vortices [6,8,9]. Filamenting pulses have been shown to propagate in atmosphere over hundred-meter to kilometer scales [10,11]. They have found applications such as fog clearing [12,13], supercontinuum generation [14], THz generation [15,16], laser-induced breakdown spectroscopy [17,18], and guided spark discharges [19,20].

Air waveguiding is one growing application of filamentation. An air waveguide is formed by a circumferential array of filaments generated by a multi-lobed beam (such as a $TEM_{11}$ Hermite-Gaussian mode) or by multi-filamentation a smooth hollow "donut" beam (such as a $LG_{01}$ Laguerre-Gaussian mode). This leads to a ring of long-lived gas density depressions, or "density holes", which thermally diffuse and merge, forming a low-index cladding surrounding the ambient air waveguide core. This fibre-like structure can guide injected optical beams [21–23] and collect remote optical signals [24,25].

A density hole arises from a single filament as follows: as a filament propagates through air, it deposits energy through plasma generation [26], excitation of molecular wavepackets [27,28], and inverse Bremsstrahlung heating [29], with the latter only relevant for multi-picosecond-long pulses. The plasma recombines and the deposited energy thermalizes within a few nanoseconds [30], which is effectively instantaneous on the acoustic response timescale of the heated region ($\tau_a \sim d_{core}/2c_s \sim 0.1\ \mu s$), where $c_s \sim 300$ m/s is the sound speed in air. This results in a pressure impulse from an initial temperature increase $\Delta T \sim 100 - 200$ K, which launches a single-cycle cylindrical acoustic wave, leaving behind a long-lived density depression or "density hole" [15,16] in pressure equilibrium with the surrounding ambient gas. The density hole lifetime can range up to tens of milliseconds as set by thermal diffusion in air [21–23,30].

In recent experiments, we demonstrated generation of and guiding in air waveguides to 50-m scales using multi-filamentation of $\sim 100$ mJ $LG_{01}$ pulses (peak intensity ring diameter $d_{ring} = 5.6$ mm and ~5 mm waveguide core diameter) at 10 Hz [22], with longer guides requiring higher pulse energy. With a thermal diffusion lifetime of ~20-30 ms, the guide dissipated between laser shots. Quasi-steady-state air waveguiding of a CW probe laser was subsequently demonstrated, with filamentation of a $TEM_{11}$ mode at > 250 Hz replenishing the ~200-μm-core-diameter waveguide faster than it could thermally dissipate [23]. More recently, using $LG_{01}$ multi-filamentation, we have demonstrated air waveguiding to $\sim 100$ m at an outdoor range [32]. Other groups have followed with $LG_{01}$-induced waveguides [33] and a demonstration of short curved air waveguides [34].

Air turbulence can have a substantial impact on filament propagation via stochastic variations in the index of refraction $n(\mathbf{r})$. The turbulence strength experienced by an optical beam is characterized by the refractive index structure parameter $C_n^2 = \langle [n(\mathbf{r}+\Delta\mathbf{r}) - n(\mathbf{r})]^2 \rangle / |\Delta\mathbf{r}|^{2/3}$, where $\Delta\mathbf{r}$ is the separation between two observation points and the angle brackets denote an ensemble average [35]. While individual near-IR filaments are smaller than the inner length scales of turbulence in air, $l_0 \sim 1$ mm [35], and therefore are primarily deflected and not broken up after collapse [36], the phase gradients induced by air turbulence can substantially impact the pre-collapse beam, affecting the collapse location [37] and inducing multi-filamentation [38] for sufficient beam energy. Multi-filamentation is induced by scintillation (fluctuating dark and 'hot' spots in the beam), the strength of which, for a laser beam of central wavenumber $k$ propagating over a distance $L$, is characterized by the Rytov variance $\sigma_R^2 = 1.23 C_n^2 k^{7/6} L^{11/6}$, where $\sigma_R^2 > \sim 1$ indicates a significant effect on the beam [39].

Up until now, air waveguiding has been demonstrated in a laser laboratory environment with measured $C_n^2 \sim 10^{-14}$ m$^{-2/3}$ [37]. However, guiding is desirable over kilometer-scale ranges in outdoor environments, where near-ground turbulence levels can range as high as $C_n^2 \sim 10^{-13}$ m$^{-2/3}$ [40]. While it was predicted in [22] that guided beams in air waveguides with 5 mm core diameter and core-cladding index contrast $\Delta n_g = n_{core} - n_{clad} \sim 10^{-6}$ (used in our ~50-m-scale guides) could robustly withstand even higher turbulence levels to $C_n^2 \sim 10^{-11}$ m$^{-2/3}$, the impact of turbulence on the air waveguide generation process itself was not considered. In this paper we demonstrate that (1) even under strong turbulence levels, $LG_{01}$ filamentation can support air waveguide generation to ~km ranges, and that (2) guided beams in air waveguides similar to [22] can withstand extreme levels of $C_n^2$ to $\sim 10^{-9}$ m$^{-2/3}$, several orders of magnitude larger than our earlier predictions.

To study air waveguiding in strong turbulence, we used the experimental setup shown in Fig. 1. An $LG_{01}$ pulse ($\lambda_0 = 812$ nm, pulse FWHM $\tau = 45$ fs, $\varepsilon_{pulse} < 100$ mJ) was down-collimated using an off-axis reflective telescope to initiate multi-filamentation around a peak intensity ring of diameter $d_{ring} = 4.2$ mm. The length of the multifilament (and air waveguide) was intentionally kept at $\sim 5$ m so that the experiment could be performed and better-diagnosed in the lab (as opposed to the 50 m hallway [22]). A $\lambda_0 = 532$ nm CW laser beam was injected into the waveguide with a $f/750$ focusing geometry through the dielectric curved mirror used for down-collimating the $LG_{01}$ beam. For all measurements, the multi-filamenting $LG_{01}$ pulse and the air-waveguided beam were directed into the outflow nozzle of a helium cell [9,41], where filamentation (and the air waveguide) terminate at the few-mm-thick air-helium interface [9] owing to $n_2^{He} / n_{2,eff}^{air} \sim 0.05$ [5,42]. This enabled in-flight imaging of the filaments and the guided beam. Waveguide efficiency was measured by imaging the guided laser mode from the air-helium interface onto a CCD camera with a 14-µs-wide electronic shutter (Fig. 1(a)). Time evolution of the guided CW pulse was measured by directing it from the interface into a photodiode-coupled integrating sphere (Fig. 1(b)). The dynamically filamenting ring of the $LG_{01}$ beam was imaged from the air-helium interface onto a CCD camera (Fig. 1(c)).

The baseline lab turbulence level was measured using deflection of the λ=532 nm CW laser, a procedure discussed in detail in Supplement 1. The standard deviation of beam centroid position $\sigma_x$ on a camera after a propagation distance $L$ is related to the atmospheric $C_n^2$ and beam diameter $D$ through $\sigma_x^2 = 0.97 C_n^2 D^{-1/3} L^3$ [43]. We measured deflection over 22 m of laboratory propagation, determining a lab turbulence level of $C_n^2 = 4 \times 10^{-14}\mathrm{m}^{-2/3}$. Additional controlled turbulence was added using a 1.8-m-long heater tape with temperature up to ~130ºC. We started the heater tape at two locations: $z = 0$ (location of the telescope's down-collimation mirror) and $z = 1$ m. The former is chosen to maximize the impact of turbulence on initial collapse and filament formation and the latter to assess its impact on sustained filament propagation and air waveguide formation. In all cases, the induced $C_n^2$ was measured using beam deflection and correlated with tape temperature, which was measured before all datasets.

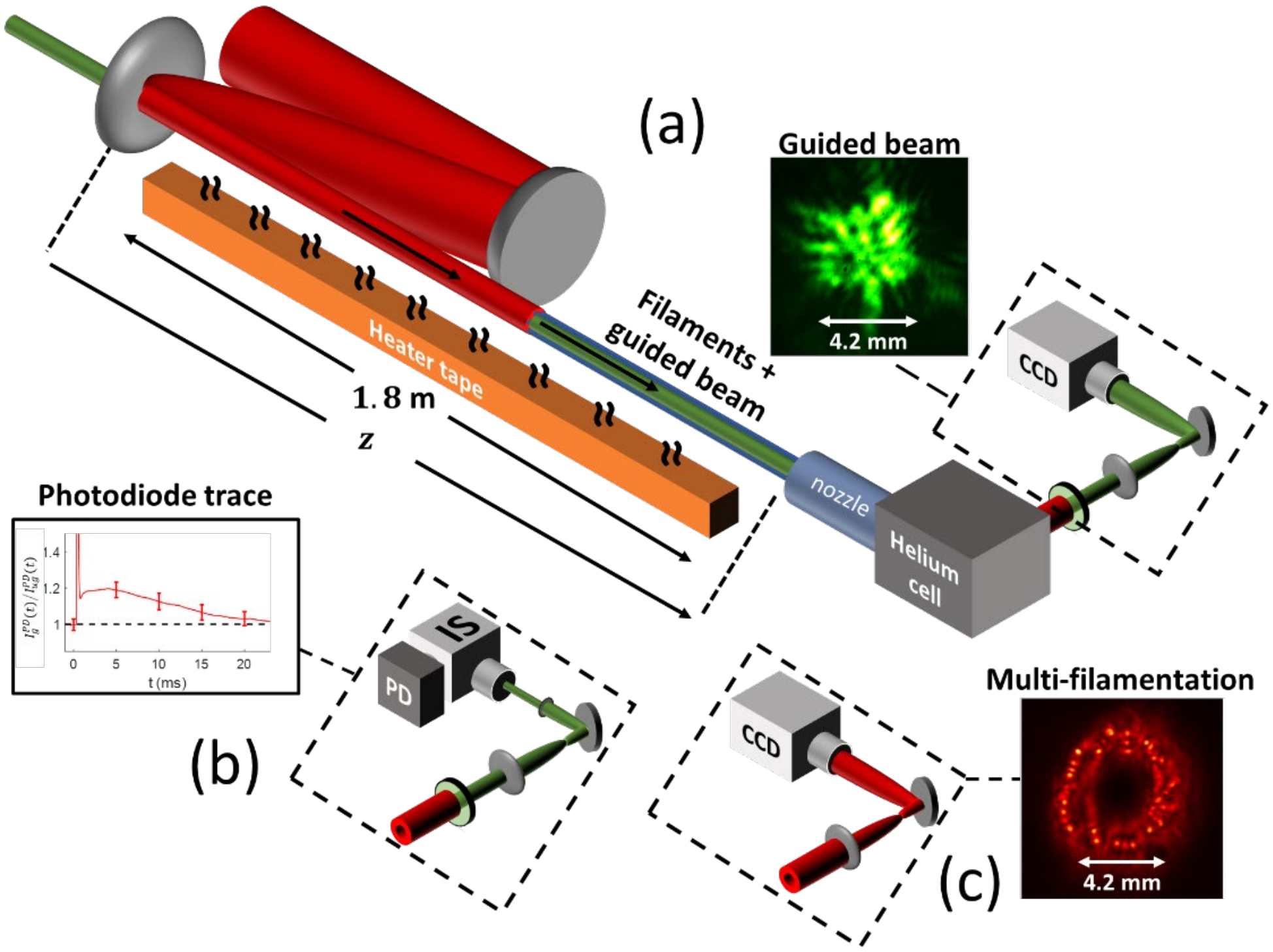

**Fig. 1.** Setup for air waveguiding through strong turbulence. **(a)** Filaments from a down-collimated $LG_{01}$ pulse propagate ~2 cm above a controlled turbulence-inducing heater tape; the helium cell is used for in-flight imaging of filaments or the guided beam onto a CCD camera, with a sample image shown of a multi-mode guided beam. This diagnostic is used to determine guiding efficiency $\eta$ (see text). **(b)** The dashed black box in (a) is replaced by a photodiode measurement, where an iris is used to isolate the core of the guided beam, which is passed through a 532 nm bandpass filter and directed into an integrating sphere (IS) coupled to a silicon photodiode (PD). This diagnostic is used to determine $I_g^{PD}(t)/I_{ug}^{PD}(t)$ (see text). **(c)** Multi-filamentation pattern at the end of the $LG_{01}$ pulse-induced filament is imaged through the helium cell (no green bandpass filter).

The filamenting beam was propagated ~2 cm above the heater tape in non-Kolmogorov turbulence, which is addressed in Supplement 1. A maximum $C_n^2 = 1.9 \times 10^{-9}$ m$^{-2/3}$ Kolmogorov-equivalent turbulence strength was measured, $\sim 200\times$ higher than the air waveguide spoiling limit of $\sim 10^{-11}$ m$^{-2/3}$estimated from [22].

## 2. Turbulence and air waveguide index structure

Results of air waveguiding of the 532 nm probe beam under varying levels of turbulence are shown in Fig. 2, where the guide is ~ 2 cm above the heater tape. Figure 2(a) shows images of the guided beam at the end of the waveguide ($z = 575$ cm) at 2 ms delay ($t = 0$ is the time of filament formation) for turbulence levels ranging over 5 orders of magnitude. In all cases, there is clear beam confinement. Row $(i)$, at lab ambient $C_n^2 = 4 \times 10^{-14}\ m^{-2/3}$, shows the characteristic multimode structure of the guided beam, where there are $V^2/2 \sim 10^3$ transverse modes, where $V \approx \frac{1}{2} k d_{ring}(NA)$ is the waveguide $V$-parameter [44], using the fact that the air waveguide is well-modeled assuming a cylindrical step index profile [22]. Here, $NA = |n_{co}^2 - n_{cl}^2|^{1/2}$ is the waveguide numerical aperture, $n_{co}$ and $n_{cl}$ are the core and cladding refractive indices (see Sec. 4) [22]. Guided multimode patterns persist at the much higher levels of turbulence of rows $(ii) - (iv)$. In row $(ii)$, guiding is demonstrated up to $C_n^2 = 1.9 \times 10^{-9}$ m$^{-2/3}$, the maximum turbulence generated by the heater tape and at least 3 orders of magnitude higher than the strongest turbulence expected outdoors near ground level [45]. Rows $(iii)$ and $(iv)$, for $C_n^2 = 0.8 \times 10^{-9}$ m$^{-2/3}$, compare the effect on air waveguiding of the heater tape (and turbulence) starting at $z = 0$, before filamentation onset, or $z = 1$ m, after filamentation is underway.

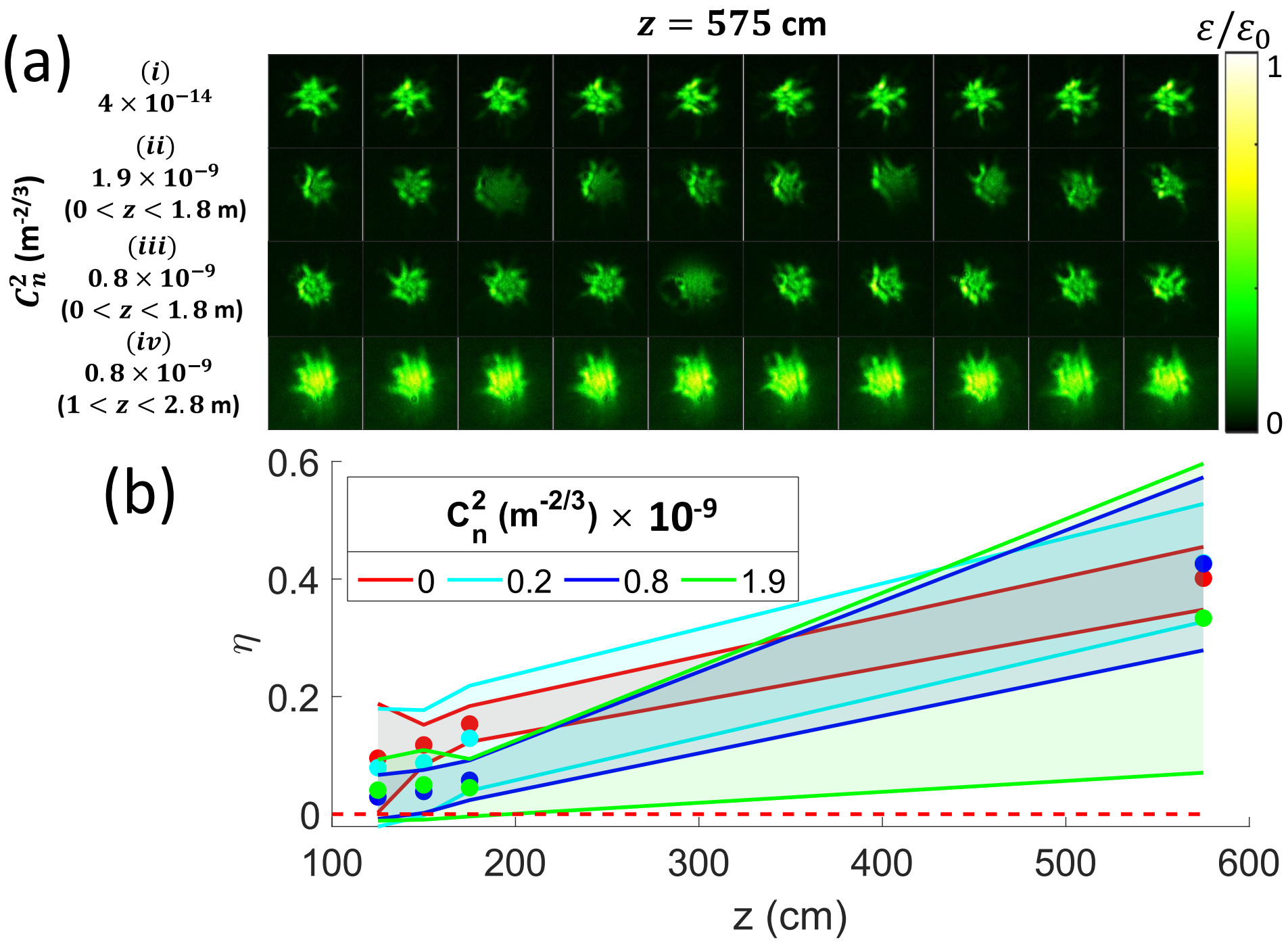

**Fig. 2. (a)** Sample images (all at 2 ms delay) of the guided beam at the end of the waveguide ($z = 575$ cm) for $(i)$ background lab turbulence level $C_n^2 = 4 \times 10^{-14}$ m$^{-2/3}$, $(ii)$ heater tape-induced turbulence $C_n^2 = 1.9 \times 10^{-9}$ m$^{-2/3}$ (heater tape $0 < z < 1.8\ m$), and heater tape-induced turbulence $C_n^2 = 0.8 \times 10^{-9}$ m$^{-2/3}$ for the tape positioned $(iii)$ $0 < z < 1.8\ m$ and $(iv)$ $1\ m < z < 2.8\ m$. The images in each row are separately normalized. **(b)** Guiding efficiency $\eta$ versus $z$ with the heater tape positioned $0 < z < 1.8\ m$. The dots plot the mean $\eta$ over 100 shots at each position; the solid lines (bounding the shaded regions) mark the $\pm$ standard deviation in $\eta$.

Figure 2(b) shows the impact of turbulence on waveguide efficiency, defined as $\eta = (E_g - E_{ug})/(E_{tot} - E_{ug})$, where $E_g$ is the guided energy in the waveguide core, $E_{ug}$ is unguided energy (no guide) within a radius equal to that of the core, and $E_{tot}$ is the total beam energy [21,22]. All are collected in a 14 $\mu s$ window starting at $t = 2$ ms. The efficiency is defined such that $\eta = 0$ for no guiding ($E_g = E_{ug}$) and $\eta = 1$ for maximum guiding ($E_g = E_{tot}$). Here, the heater tape occupies $0 < z < 1.8\ m$. Guiding measurements were performed over heater tape and at the end of the guide, with the average $\eta$ (solid circles) taken over 100 shots for each position. The solid curves linearly link the points at the end of the tape and those at the end of the waveguide, marking the $\pm$ standard deviation of the $\eta$ measurements. It is seen that $\eta$ slightly decreases with increasing $C_n^2$ in the region over the heater tape, where the beam is still coupling into the guide. The standard deviation in $\eta$ also increases with $C_n^2$, indicating a large shot-to-shot variation in waveguide performance at maximum turbulence strength. This is also demonstrated in the guided shots shown in Fig. 2(a), where, for example, shot 3 in row $(ii)$ shows worse guiding than shot 10.

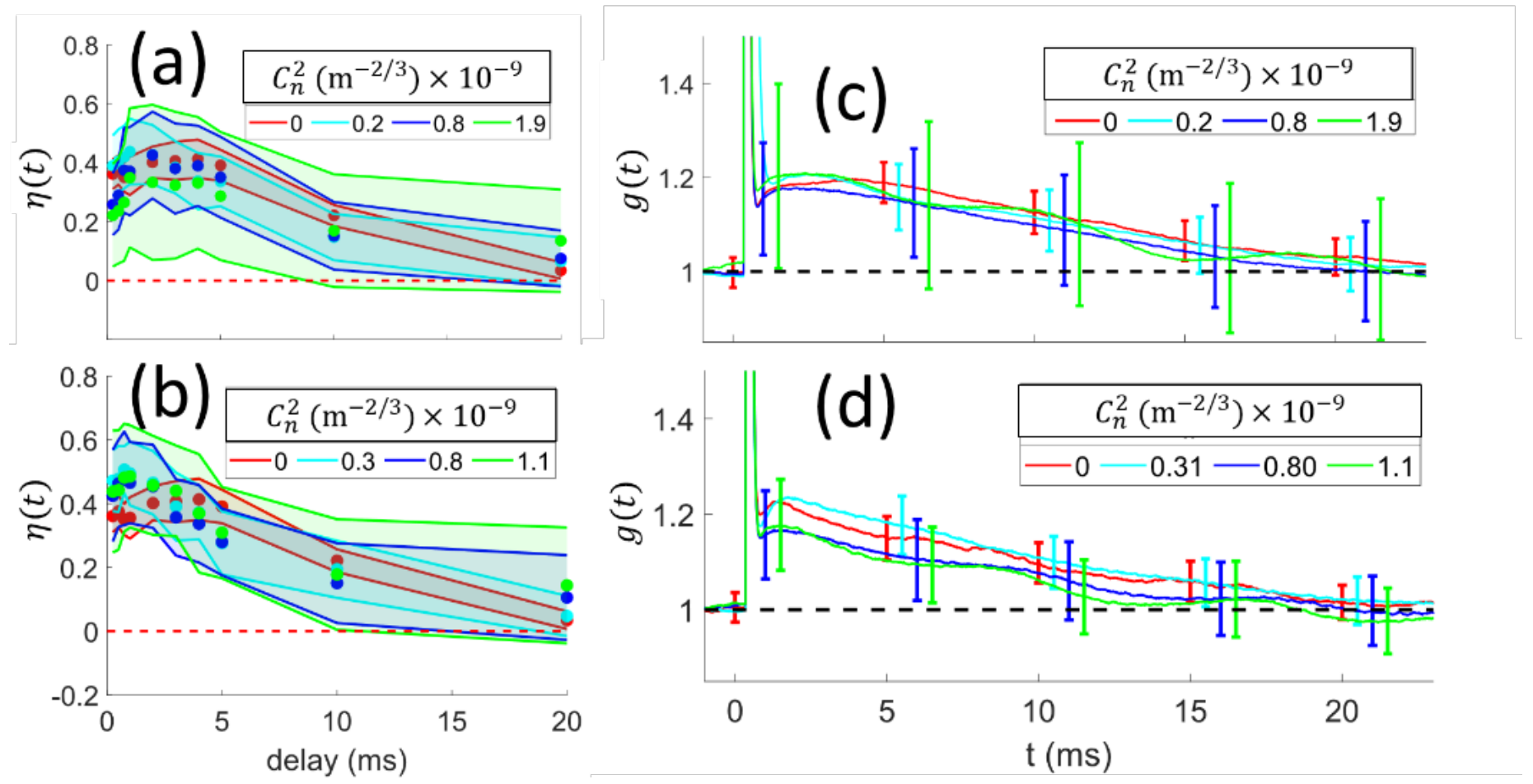

**Fig. 3.** Waveguide performance over time at z = 575 cm. **(a)** Time-dependent guiding efficiency $\eta(t)$ with the heater tape located at $0 < z < 1.8\ m$. The solid lines (and enclosed shading) follow. the $\pm$ standard deviation. **(b)** Efficiency $\eta(t)$ with the heater tape located at $1\ m < z < 2.8\ m$. **(c)** Time-dependent guided pulse enhancement $g(t) = I_g^{PD}(t)/I_{ug}^{PD}(t)$ with tape located at $0 < z < 1.8\ m$ **(d)** $g(t)$ traces with the heater tape located at $1\ m < z < 2.8\ m$.

Figure 3(a) plots time-dependent waveguide efficiency $\eta(t)$ at $z = 575$ cm, the end of the guide, where $\eta(t) = (E_g(t) - E_{ug}(t))/(E_{tot}(t) - E_{ug}(t))$, with the energies collected at time $t$ with a 14 $\mu s$ electronic camera window. Here the heater tape was located $0 < z < 1.8\ m$. Figure 3(b) plots the same thing for the heater tape located at $1\ m < z < 2.8\ m$. As in Fig. 2, the points are 100 shot averages, with the solid lines (and enclosed shading) following the $\pm$ standard deviation. Figures 3(c) and (d) plot the signal enhancement $g(t) = I_g^{PD}(t)/I_{ug}^{PD}(t)$, the ratio of the guided and unguided photodiode signal (see Fig. 1(b)), for the heater tape at $0 < z < 1.8\ m$ and $1$ m $< z < 2.8$ m respectively. Error bars, corresponding to the $\pm$ standard deviation at each point, are overlaid on the curves. The photodiode, unlike the CCD camera, could not simultaneously collect the core and total beam energies ($E_g$ and $E_{tot}$); $g(t)$ was therefore chosen as its associated measure of guiding enhancement.

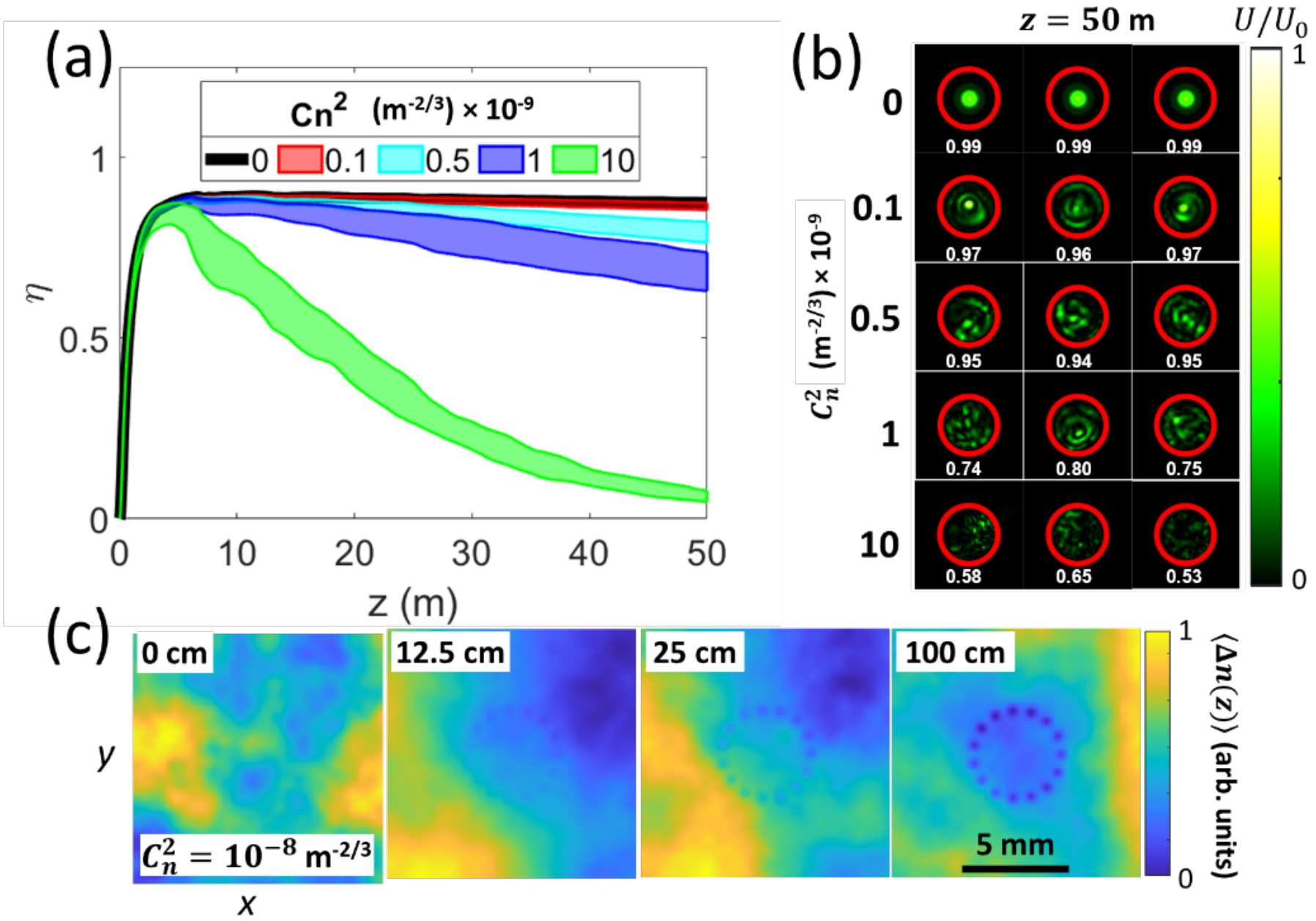

**Fig. 4.** Propagation simulations demonstrating impact of turbulence on air waveguiding for a unform density hole distribution. **(a)** Guiding efficiency $\eta$ vs. distance for increasing $C_n^2$, 2 ms delay and $d_{guide} = 4.2\ mm$. **(b)** Sample mode structure at $z =$ 50 m for simulations with three sets of random phase screens (columns) at five $C_n^2$ levels (rows). The waveguide core is indicated by the red circle, with the fraction of energy in the core labeled in each image. **(c)** Axial average index of refraction $\langle \Delta n(x,y) \rangle_z$ for $C_n^2 = 10^{-8}$ m$^{-2/3}$ up to $z = 100$ cm. The color bar limits for each image are the minimum and maximum values of $\langle \Delta n(z) \rangle$ to highlight density hole structure against the turbulent background.

A striking observation is that the shot-averaged $\eta$-traces, as well as the $g$-traces, are comparable in amplitude and temporal variation in spite of the turbulence strength varying over ~5 orders of magnitude. However, standard deviation from the averages increases with turbulence strength. The evolution of the $g(t)$ traces in Figs. 3(c) and 3(d) generally matches those of $\eta(t)$ in 3(a) and 3(b), with some variation due to changes in waveguide coupling between datasets. The initial spike in the $g(t)$ curves is due to filament-generated supercontinuum light leaking through the 532 nm bandpass filter (See Fig. 1(b)) and saturating the photodiode. After this spike, the trace follows the time-dependent guided intensity.

Taken together, the results in Figs. 2 and 3 show that air waveguides operate far beyond the predicted $C_n^2$ threshold of $\sim\Delta n_g^2/d_{ring}^{2/3} \sim 10^{-11}$ m$^{-2/3}$ based on our experimental parameters [22]. We explain this behavior using beam propagation method (BPM) simulations [46] of linear propagation over 50 m, where the turbulence is modeled using random phase screens (see Supplement 1). Figure 4(a) shows simulated guide efficiency $\eta$ for $C_n^2$ values up to $10^{-8}$ m$^{-2/3}$. In all cases there is guiding, although the performance degrades strongly for $C_n^2 > \sim 10^{-9}$ m$^{-2/3}$, which will be explained in Sec. 4. Over the first several meters, however, guiding is efficient for all turbulence strengths sampled. Figure 4(b) shows, for five levels of turbulence ranging up to $C_n^2 = 10^{-8}$ m$^{-2/3}$, three sample guided modes resulting from random phase screens associated with each level of turbulence. The core region is marked by a red circle and the fraction of guided energy in the core is indicated on each image. As $C_n^2$ increases, the shot-to-shot variations in mode structure clearly increase. However, even for the strongly perturbed beams at high $C_n^2$, energy confinement is still high.

An explanation for this result is found in Fig. 4(c), which plots the axially averaged index of refraction $\langle \Delta n(x,y)\rangle_z = \frac{1}{z}\int_0^z \Delta n(x,y,z')dz'$ from $z = 0$ cm to $z = 100$ cm for $C_n^2 = 10^{-8}$ m$^{-2/3}$. At $z = 0$ ($\langle \Delta n(x,y)\rangle_{z=0}$), the guiding index structure is entirely obscured by the phase screen, but it emerges for $z > 12.5$ cm as the effects of the turbulent index variations average to zero. *Thus, barring the scattering of the guided beam out of the waveguide by strong index fluctuations, turbulent index variations orders of magnitude larger than the core-cladding index difference will not impede guiding.* By contrast, an unconfined Gaussian beam experiences no restoring index structure, so each turbulent phase shift or angular perturbation alters its subsequent free propagation; the accumulated second order index variances manifest as beam wander, spreading, and scintillation.

The rest of this manuscript focuses on the two other deleterious effects of turbulence: (1) its impact on $LG_{01}$ multi-filamentation and waveguide formation, and (2) the aforementioned scattering of the guided beam out of the waveguide.

## 3. $LG_{01}$ multi-filamentation in turbulence

We now address the effect of turbulence on $LG_{01}$ multi-filamentation and air waveguide cladding formation. A straightforward measure of beam distortion is the scintillation index $\sigma_I^2(x,y) = \langle I^2(x,y)\rangle_z / \langle I(x,y)\rangle_z^2 - 1$ [39], where $\langle I\rangle_z$ and $\langle I^2\rangle_z$ are the measured average intensity and squared intensity of the beam over $N = 500$ pulses at axial location $z$. Given that filaments form near the radial location of peak intensity, the level of scintillation in that region is especially important for assessing the effect of turbulence on air waveguide formation. Accordingly, we define the peak intensity scintillation index as $\sigma_I^2(r_p) = \langle I^2(r_p)\rangle_z / \langle I(r_p)\rangle_z^2 - 1$, where $r = r_p$ is the radial location of peak intensity of the $N$-shot image average $\langle I(x,y)\rangle_z$, $\langle I(r_p)\rangle_z$ is the $N$ −shot average peak intensity averaged around the ring $r = r_p$, and $\langle I^2(r_p)\rangle_z$ is the corresponding average squared peak intensity. In computing $\sigma_I^2(r_p)$ from experimental data, we effectively sample a band $r_p \pm \Delta r_p$, where $\Delta r_p \sim 4$ μm is set by the image resolution. Figure 5(a) shows three in-flight experimental images of the filamenting beam for each of three turbulence strengths. In this data set, the heater tape occupied $0 < z < 1.8\ m$, with the images taken at $z = 3.2$ m. The only notable shot-to-shot variation occurs for $C_n^2 = 1.9 \times 10^{-9}$ m$^{-2/3}$, where it is seen that the filament ring is still maintained and the waveguide cladding can still form.

Figure 5(b)(i) plots $\sigma_I^2(r_p)$ for $LG_{01}$ pulses propagating through $C_n^2 \sim 0$ (room turbulence $4 \times 10^{-14}$ m$^{-2/3}$, blue curves and points) and $C_n^2 = 1.9 \times 10^{-9}$ m$^{-2/3}$ (red curves and points) for low power (5 mJ, pulsewidth 10 ns, solid circles) and high-power (100 mJ, 45 fs, stars). The solid curves are from BPM simulations through turbulence. First, the points for low and high-power pulses (circles and stars) largely track each other for both low and high turbulence cases, indicating that nonlinear propagation has little effect on shot-to-shot scintillation variations. The experimental points are overlaid by linear BPM simulations of propagation assuming Kolmogorov turbulence with a power spectral density $\Phi_n(\kappa) \propto \kappa^{-11/3}$, where $\kappa$ is the turbulence wavenumber [47]. While the low turbulence simulation (blue solid curve) matches experiment, the high turbulence simulation (red solid partial curve) diverges substantially from it. This is due to the turbulence being non-Kolmogorov over our propagation path (as measured in Supplement 1), with the estimated spectrum $\Phi_n(\kappa) \sim \kappa^{-11.7/3}$, producing weaker scintillation [48–50]. Repeating the BPM simulation using this turbulence spectrum gives the dashed red curve, which better overlaps the data points. Figure 5(b)(ii) is a full-scale plot comparing the $\kappa^{-11/3}$ and $\kappa^{-11.7/3}$ simulations with the high power, high turbulence data (red stars). It is worth noting that the scintillation index $\sigma_I^2(r_p)$ is substantially lower than the Rytov variance $\sigma_R^2$ for both our simulations and experiments due to non-plane-wave effects and non-Kolmogorov turbulence. For the experimental data (and non-Kolmogorov turbulence simulation), $\sigma_R^2/\sigma_I^2(r_p) \sim 22$, and for the simulation with Kolmogorov turbulence, $\sigma_R^2/\sigma_I^2(r_p) \sim 3$. This indicates that a waveguide structure can be formed in spite of high levels of Rytov variance.

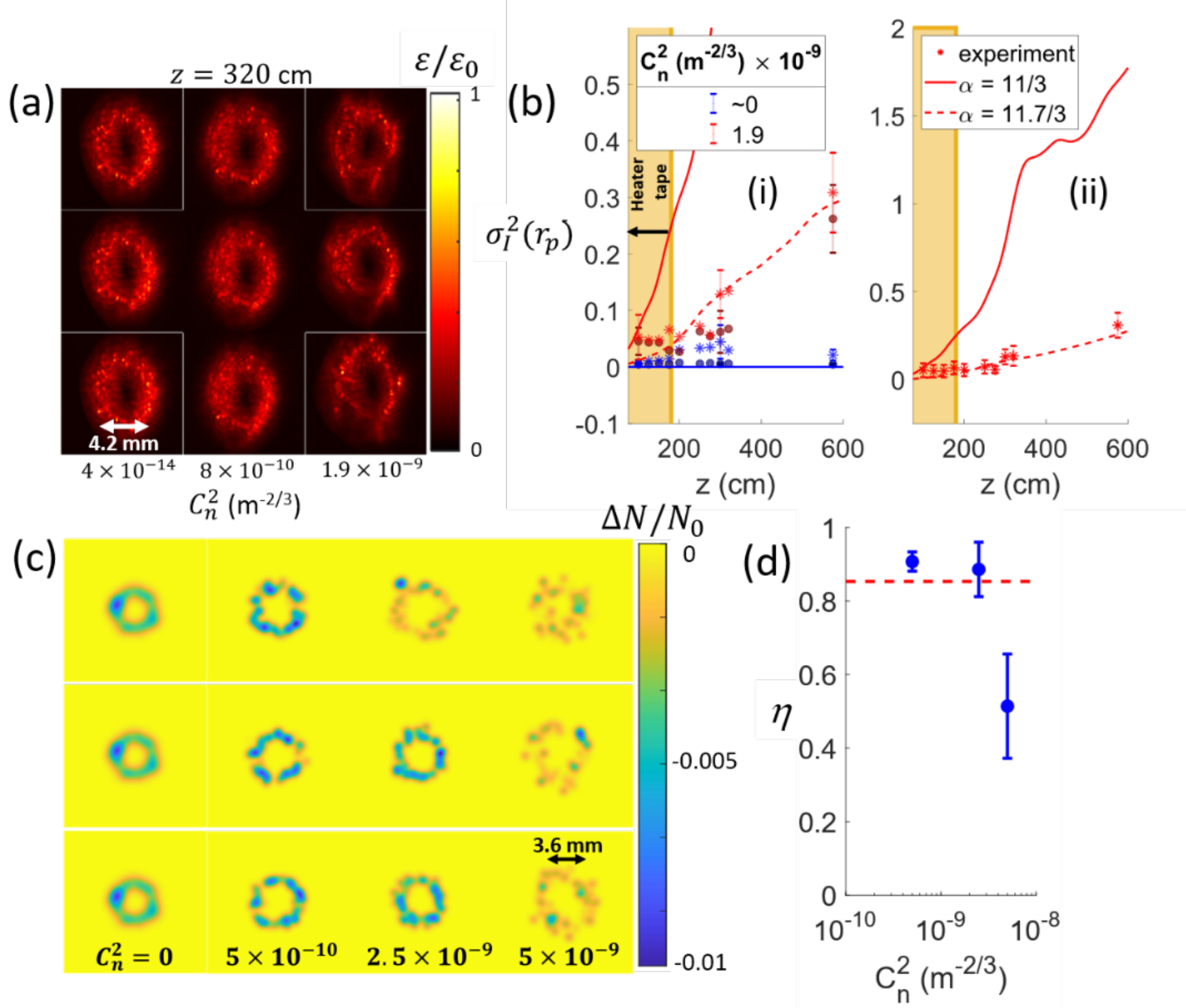

**Fig. 5.** **(a)** In-flight filament images from helium cell at $z = 320$ cm, with the heater tape occupying $0 < z < 1.8\ m$. The columns are three sample shots at each turbulence strength. **(b)** $(i)$ Peak-intensity scintillation index $\sigma_I^2(r_p)$ vs. propagation distance for $C_n^2 \sim 0$ (blue points and curves) and $C_n^2 = 1.9 \times 10^{-9}$ m$^{-3/2}$ (red points and curves). Solid circles: low power (5 mJ, pulsewidth 10 ns); stars: high power (100 mJ, 45 fs). Blue and red solid curves are from BPM simulations of these low and high $C_n^2$ cases assuming Kolmogorov turbulence ($\propto \kappa^{-11/3}$). Red dashed curve: high $C_n^2$ turbulence using non-Kolmogorov turbulence ($\propto \kappa^{-11.7/3}$). Selected error bars correspond to $\pm$ standard deviation over $N = 500$ shots. $(ii)$ Full scale plot of high power, high turbulence data and curves of (i). **(c)** Nonlinear propagation simulations using YAPPE (see Supplement 1): Three runs to $z = 1.6$ m for each of four $C_n^2$ levels simulating density-hole generation. **(d)** Simulated waveguide efficiency $\eta$ over 50 m using the density hole profiles in (c). The dashed red line plots baseline efficiency for $C_n^2 = 0$, using the density hole cladding profiles in the leftmost column of (c).

The impact of turbulence on filament ring and density hole formation was examined using our propagation code YAPPE, an implementation of the unidirectional pulse propagation equation (UPPE) [6,51–53], to simulate nonlinear optical propagation through Kolmogorov turbulence. Using a smaller diameter guide (limited by computer memory), Fig. 5(c) shows density hole profiles resulting from three sample propagation simulations for an $LG_{01}$ pulse (75 mJ, $\tau = 40$ fs, $d_{ring} = 3.6$ mm) for each of four turbulence levels up to $C_n^2 \sim 5 \times 10^{-9}$ m$^{-2/3}$. The density hole profiles are calculated by time-integrating the local power absorption from the filament, converting the absorption profile to a temperature (and associated density) profile of neutral air, and then letting it thermally diffuse for 2 ms. See Supplement 1 for details of this calculation. These results demonstrate that significant filament ring (and density hole) variation is not seen until $C_n^2 \sim 5 \times 10^{-9}$ m$^{-2/3}$, corresponding to $\sigma_R^2 \sim 1.6$ (for $L = 1.6$ m). In our simulated Kolmogorov turbulence, the scaling $\sigma_R^2 \sim 3\sigma_I^2(r_p)$ implies $\sigma_I^2(r_p) \sim 0.5$ for the onset of cladding degradation. This is further supported by using the density hole structures in Fig. 5(c) as the waveguide cladding in BPM simulations, with efficiency $\eta$ plotted vs. $C_n^2$ in Fig. 5(d). The plot shows that waveguide performance does not degrade over $\sim 2$ m until $C_n^2 \sim 5 \times 10^{-9}$ m$^{-2/3}$, or as $\sigma_I^2(r_p)$ approaches 1. Using a $\sigma_I^2 \sim L^{11/6}$ scaling, we can expect successful waveguide formation for $C_n^2 < 2 \times 10^{-12}$ m$^{-2/3}$ over 100 m and $C_n^2 < 3 \times 10^{-14}$ m$^{-2/3}$ over 1 km.

## 4. Scattering losses from turbulence

The other main effect impacting air waveguide performance is possible turbulent scattering of the guided beam out of the guide. As the air waveguide is well-modeled assuming a cylindrical step-index profile, its numerical aperture is $NA = \sqrt{n_{co}^2 - n_{cl}^2} \approx (2n_0|\delta n_{cl}|)^{1/2}$, where we use $n_{co} = n_0 + \delta n_{co}$, $n_{cl} = n_0 + \delta n_{cl}$, $\delta n_{co} = 0$, and $|\delta n_{cl}/n_0| \ll 1$, where $\delta n_{cl}$ is the cladding index decrement from ambient $n_0$ [22]. Propagating rays are guided for $\theta < \theta_{max}$, where $\theta$ is the ray approach angle with respect to the guide axis and $\theta_{max} = \sin^{-1}(NA)$ is the critical acceptance angle of the waveguide. For a typical core-cladding index contrast ($\Delta n_g = n_{co} - n_{cl} \approx |\delta n_{cl}| = 10^{-6}$ [22]), $NA \sim 1.4 \times 10^{-3}$. Turbulent eddies attenuate the beam by scattering it into angles $\theta > \theta_{max}$.

We determine the effect of turbulence by using the Mutual Coherence Function (MCF) [39] to calculate the scattering angular distribution, whose integral over the acceptance aperture ($0 < \theta < \theta_{max}$) gives the relation between the fraction of beam energy $f_{g,turb}(L)$ supported by the guide at $z = L$ and that in the guide region without the waveguide present, $f_{ug,turb}(L)$ (see Supplement 1):

$$f_{g,turb}(L) = f_{ug,turb}(L) + \left(1 - f_{ug,turb}(L)\right) \operatorname{erf}\left[\frac{0.497\, \lambda^{1/5} \theta_{max}}{(C_n^2 L)^{3/5}}\right]. \quad (1)$$

Using Eq (1), the effective waveguide efficiency is $\eta_{turb}(L) = \left[f_{g,turb}(L) - f_{ug,turb}(L)\right] / [1 - f_{ug,turb}(L)]$, which becomes, for Kolmogorov turbulence,

$$\eta_{turb}(L) = \operatorname{erf}\left[\frac{0.497\, \lambda^{1/5} \theta_{max}}{(C_n^2 L)^{3/5}}\right] \quad (2)$$

In Fig. 6(a), $\eta_{turb}(L)$ is plotted for $L < 1$ km for several values of $C_n^2$. It is seen that for $C_n^2 \sim 10^{-10}$ m, turbulence substantially scatters light out of the guide by 1 km. For increasing values of $C_n^2$, scattering losses occur earlier. At $C_n^2 \sim 10^{-8}$ m$^{-2/3}$, the guide hardly confines the beam beyond $\sim 10$ m. These efficiency curves qualitatively follow the simulation curves of Fig. 4(a), where guiding over 50 m falls off substantially between $C_n^2 \sim 10^{-9} - 10^{-8}$ m$^{-2/3}$.

Using a similar analysis, one can determine the turbulence threshold for significant guided beam scattering; this is where the acceptance angle of the waveguide does not contain a large portion of the beam. By setting the $1/e^2$ width of the MCF angular distribution equal to $\theta_{max}$ (see Supplement 1), the threshold for significant turbulent scattering is

$$(C_n^2 L)_{thresh} \sim \frac{\lambda^{1/3}}{5.72} \left(\sin^{-1}\left[(n_{co}^2 - n_{cl}^2)^{1/2}\right]\right)^{5/3}, \quad (3)$$

which depends only on the waveguide numerical aperture and the wavelength of the guided light. A log-log plot of threshold $C_n^2$ versus $L$ is shown in Fig. 6(b) for a range of core-cladding index contrasts $\Delta n_g$. For reference, $\Delta n_g = 10^{-6}$ is the estimated core-cladding index contrast for a typical air waveguide at 2 ms delay [22]. As applied to our experiments, the plot indicates that scattering by turbulence will not affect guiding over $\Delta z = 1.8$ m until $C_n^2 \sim 10^{-8}$ m$^{-2/3}$, a level of turbulence we could not reach experimentally.

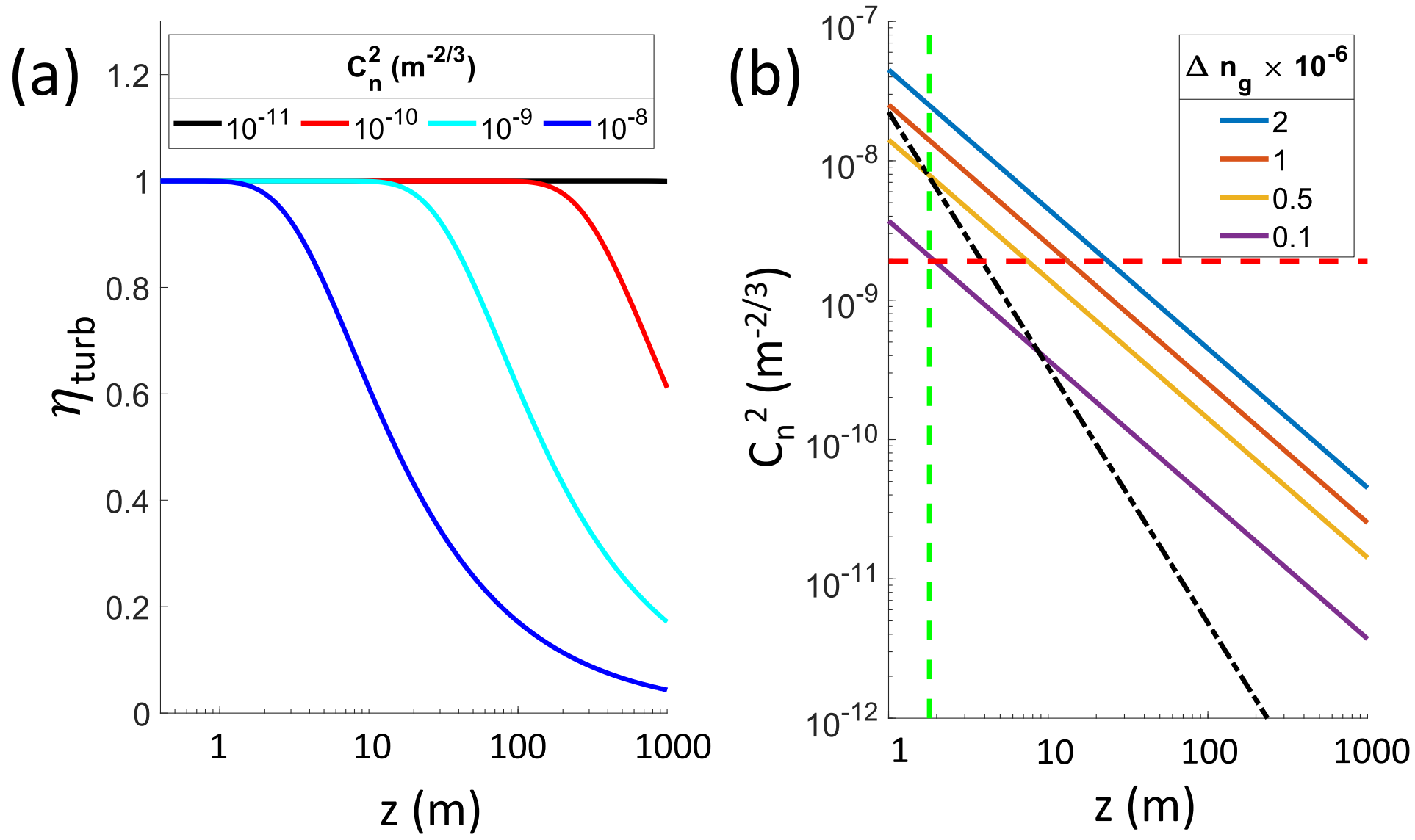

**Fig. 6.** **(a)** Calculated efficiency $\eta_{turb}$ vs. z for a step index waveguide in various levels of turbulence $C_n^2$, $\Delta n_g = 10^{-6}$. **(b)** Calculated $C_n^2$ threshold for significant scattering by turbulence for various values of the core-cladding index $\Delta n_g$ as a function of distance $z$. The red horizontal dashed line marks $C_n^2 = 1.9 \times 10^{-9}$ m$^{-2/3}$ (maximum experimental turbulence), and the green vertical dashed line marks $z = 1.8$ m (the length of the heater tape). The black dashed line corresponds to $\sigma_I^2 \sim 1$ in Kolmogorov turbulence assuming $\sigma_I^2 \sim \sigma_R^2/3$, as discussed in Sec. 3.

## 5. Conclusions

We have considered three deleterious effects of turbulence on air waveguiding: (1) loss of waveguide confinement because of index fluctuations exceeding the core-cladding index contrast, (2) scattering of guided light out of the waveguide by sufficiently large index fluctuations, and (3) the effect of turbulence-induced scintillation on long range filament ring formation and air cladding generation. Our results establish turbulence limits for long-range optical guiding in the atmosphere.

We found that turbulence-induced index fluctuations orders of magnitude larger than the air waveguide index contrast leave beam confinement unaffected owing to the axial averaging of the fluctuations experienced by the guided beam. As a result, air waveguiding can occur in far higher levels of turbulence than originally predicted [22]. Likewise, the index fluctuations required to scatter light out of a typical air waveguide are orders of magnitude larger than those typically found in the lab or in outdoor environments. By contrast, a freely propagating Gaussian beam experiences no guiding structure and is subject to accumulated second order index variances that cause beam wander, spreading, and scintillation.

The most deleterious effect of turbulence is scintillation of the waveguide generation beam, which can degrade long range filament ring and waveguide cladding formation. Even there, our experiments, simulations, and parameter scalings indicate that highly confining 1-km-long air waveguides can be generated in Kolmogorov turbulence levels as high as $C_n^2 \sim 10^{-14} - 10^{-13}\ \mathrm{m}^{-2/3}$, a strong level of turbulence near ground level. Finally, we note that because scintillation decreases with increasing wavelength, longer-wavelength filament drivers may further extend the accessible turbulence range.

**Funding.** This work was supported by the Air Force Office of Scientific Research (FA9550-25-1-0332), the Office of Naval Research (N00014-20-1-2233) and the Department of Energy (DE-SC0024398).

**Acknowledgements.** The authors thank Scott Hancock and Lucas Railing for technical discussions.

**Disclosures.** The authors declare no conflicts of interest.

**Data availability.** Data underlying the results presented in this paper are not publicly available at this time but may be obtained from the authors upon reasonable request.

**Supplemental document.** See Supplement 1 for supporting content.

# Optical air waveguides in strong turbulence: supplemental material

A. GOFFIN[1,2], G.I. BABIĆ[2], AND H.M. MILCHBERG[2*]

[1]Physical Chemistry and Applied Spectroscopy, Los Alamos National Laboratory, Los Alamos, NewMexico 87545, USA
[2]Institute for Research in Electronics and Applied Physics, University of Maryland, College Park, Maryland 20742, USA
*milch@umd.edu

## 1. Beam deflection measurements of $C_n^2$

The values of refractive index structure parameter $C_n^2$ reported in this work were determined by measuring the deflection of a $\lambda = 532$ CW nm (green) laser beam propagating over the heater tape on the same path as the filamenting beam, with the green beam imaged using a CCD camera. Data sets were collected with the heater tape in two configurations: (1) placed on thick aluminum blocks to thermally separate the heater tape from the optical table and (2) suspended tautly above the optical table using thin posts. In configuration (1), voltages 30V, 60V, 90V, and 120V were applied to the heater tape, corresponding to a temperature range 75°F-170°F. In configuration (2), voltages 90V-140V in 10V increments were applied, corresponding to 160°F - 270°F. The green beam traveled over a total path $l = 4.84$ m, the first $L = 1.82$ m of which was over the heater tape, and a set of 1000 images of the beam (each with a 14-µs-wide electronic shutter) were collected for each temperature setting. The intensity centroid was calculated for each image, along with the centroid mean and standard deviation $\sigma_c$ for each set of 1000 images; the tangent of the beam deflection angle is $\tan\theta_d = \sigma_c/l \approx \theta_d$. The $C_n^2$ values were then calculated using $C_n^2 L = \theta_d^2 D^{1/3}/0.97$ [1] where $D = 2$ mm is the green beam's transverse FWHM. This gives

$$C_n^2 = \frac{\sigma_c^2 D^{1/3}}{0.97 L l^2} \ . \tag{S1}$$

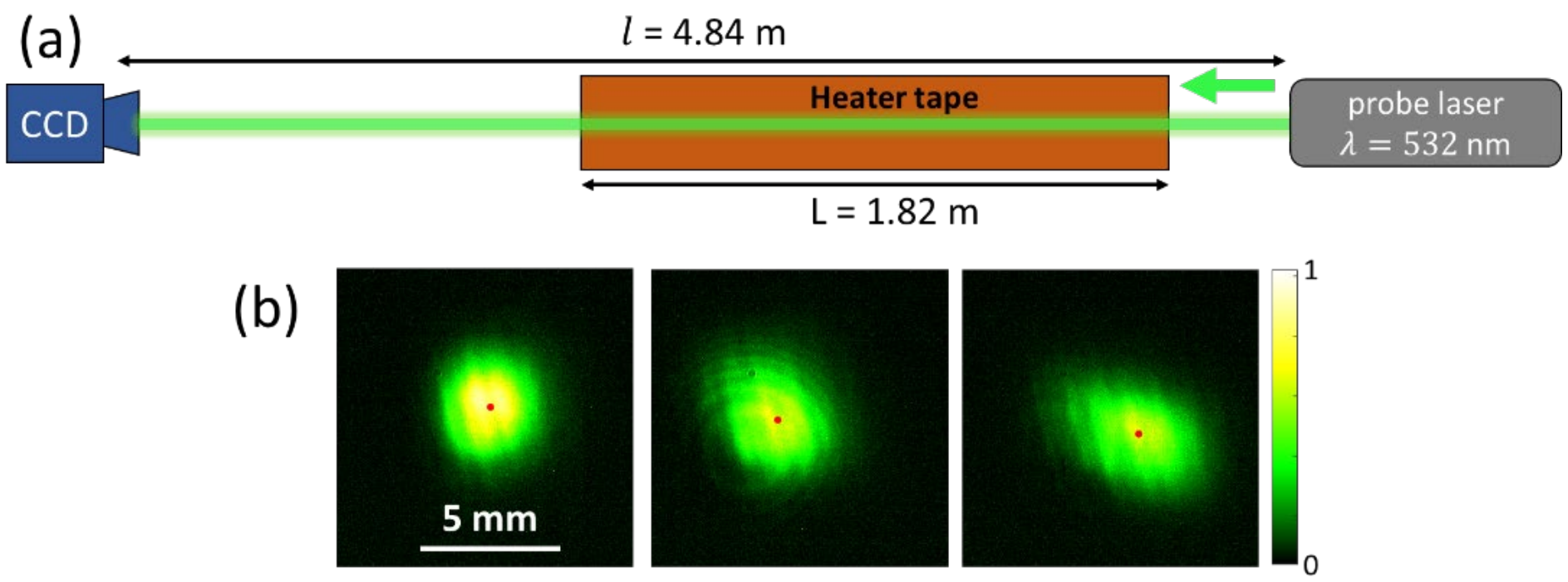

**Figure S1. (a)** Diagram of green laser deflection measurements, showing the length of the heater tape $L$ and length of propagation $l$. **(b)** Three sample shots at the maximum measured $C_n^2 = 1.9 \times 10^{-9}$ m$^{-2/3}$. The red dots mark the calculated centroid for each shot.

Using Eq. (S1), we determined $C_n^2$ for each tape configuration and voltage, with $C_n^2$ ranging from $\sim 10^{-10}$ m$^{-2/3}$ to $1.9 \times 10^{-9}$ m$^{-2/3}$. Deflection measurements were also performed with no heater tape over 22 m of lab propagation to measure lab turbulence. In this case, $L = l = 22$ m and equation (S1) yielded $C_n^2 = 4 \times 10^{-14}$ m$^{-2/3}$ (as reported in the main text).

## 2. Laser propagation simulations

To simulate linear, monochromatic beam propagation in this work, we used the Beam Propagation Method (BPM) [2], a split-step Fourier-based paraxial solver. It consists of a diffractive step (applying a Fraunhofer diffraction term $e^{ik_\perp^2/(2k_0)\delta z}$ to radial wavenumbers $k_\perp$) followed by a refractive step (applying a refractive index shift $e^{-ik_0 \Delta n \delta z}$) with simulation steps $\delta z$ in the propagation direction. Turbulence is applied in the refractive step via refractive index screens $\Delta n_{turb}(x, y; z)$ [3]. "Kolmogorov" index screens, as they are referred to in the main text, are generated based on a modified von Karman power spectral density [4]:

$$\Phi_{MVK}(\kappa) = 0.033 C_n^2 (\kappa^2 + \kappa_0^2)^{-11/6} e^{-\kappa^2/\kappa_m^2} \tag{S2}$$

where $\kappa_0 = 2\pi/L_0$ is based on the outer scale length of turbulence $L_0 \sim 1$ m (height from the ground) [5], $\kappa_m = 5.92/l_0$ is based on the inner scale length of turbulence $l_0 \sim 1$ mm [5], and $C_n^2$ is the refractive index structure parameter. Phase screens are generated by multiplying this spectrum by a normally distributed set of complex numbers in 3D-space then taking the real part of the inverse 3D Fourier transform.

For non-Kolmogorov turbulence, the power spectral density follows [6]

$$\Phi_{NK}(\kappa) = A(\alpha) C_n^2 L_0^{11/3-\alpha} (\kappa^2 + \kappa_0^2)^{-\alpha/2} e^{-\kappa^2/\kappa_m^2}, \tag{S3}$$

where $A(\alpha) = (4\pi^2)^{-1}\Gamma(\alpha - 1)\cos(\alpha\pi/2)$ [7] and $\alpha$ is the characteristic exponent of the turbulence inertial range (measured in Sec. 3 below). Note that for $\alpha = 11/3$, Eq. (S3) is equivalent to Eq. (S2). For both three-dimensional power spectra (in general, $\Phi_n(\kappa)$), $\Delta n_{turb}(x, y; z)$ is generated with a complex-valued random seed $r$ of modulus one through the 3D Fourier transform $\Delta n_{turb}(x, y; z) = \mathcal{F}_{\kappa_x,\kappa_y,\kappa_z}\{\sqrt{r\Phi_n(\kappa)}\}$.

Nonlinear propagation through turbulence was simulated using YAPPE [8–10], an implementation of the unidirectional pulse propagation equation (UPPE) [11], in conjunction with turbulent phase screens. YAPPE simulates nonlinear pulse propagation using

$$\partial_z A_{k_x,k_y}(\omega; z) = i2\pi Q_{k_x,k_y}(\omega) P_{k_x,k_y}(\omega; z) e^{-i\left(k_z - \frac{\omega}{v_g(\omega_0)}\right)z}, \tag{S4}$$

where $A_{k_x,k_y}(\omega; z)$ is the 3D inverse Fourier transform of the z-shifted electric field $E(x, y, \tau; z)$: $A_{k_x,k_y}(\omega; z) = \mathcal{F}^{-1}_{x,y,\tau}\{E(x, y, \tau; z)e^{-ik_z\Delta z}\}$. In YAPPE, by contrast with BPM, the laser field is an ultrashort (and therefore polychromatic) pulse. In Eq. (S4) and in the definition of $A$, $\tau = t - z/v_g(\omega_0)$ is shifted time in a frame moving at the group velocity $v_g(\omega_0)$, $\omega_0$ is the central angular frequency, $\Delta z$ is the simulation step size, $k_z = [(\omega/v_g(\omega_0))^2 - (k_x^2 + k_y^2)]^{1/2}$ is the longitudinal spatial frequency, and $Q_{k_x,k_y}(\omega) = \omega/ck_z$. The transverse spatial frequencies $(k_x, k_y)$ index a system of ordinary differential equations, with one ODE for each pair of spatial frequencies. $P_{k_x,k_y}(\omega; z)$ is the nonlinear polarization of the medium, including Kerr self-focusing, rotational nonlinear self-focusing [12], ionization dynamics [13], and a dispersive plasma response. At each simulation step, turbulence is applied via a temporally-constant phase screen $\Delta\phi_{turb}(x, y; z)$ to the auxiliary field $A(x, y, \tau; z) = \mathcal{F}_{k_x,k_y,\omega}\left\{A_{k_x,k_y}(\omega; z)\right\} e^{i\Delta\phi_{turb}(x,y;z)}$. $\Delta\phi_{turb}(x, y; z)$ is from $\Delta\phi_{turb}(x, y; z) = k_0 \Delta n_{turb}(x, y; z)\Delta z$, where $\Delta n_{turb}(x, y; z)$ is generated from the turbulence power spectrum as discussed above.

### 3. Measurement of turbulence spectrum

To correctly analyze behavior related to turbulent scattering and scintillation, it was essential to measure the turbulence spectrum over the heater tape. In standard analysis of optical propagation through turbulence, atmospheric spectra are of Kolmogorov form, $\Phi_K(\kappa){\sim}\kappa^{-11/3}$ in 3D space (neglecting inner and outer scales). However, in general, turbulence can be non-Kolmogorov with a form $\Phi_{NK}(\kappa){\sim}\kappa^{-\alpha}$ for some $\alpha$. Specific forms of these spectra based on the modified von Karman spectrum are shown in Eqs. (S2) and (S3).

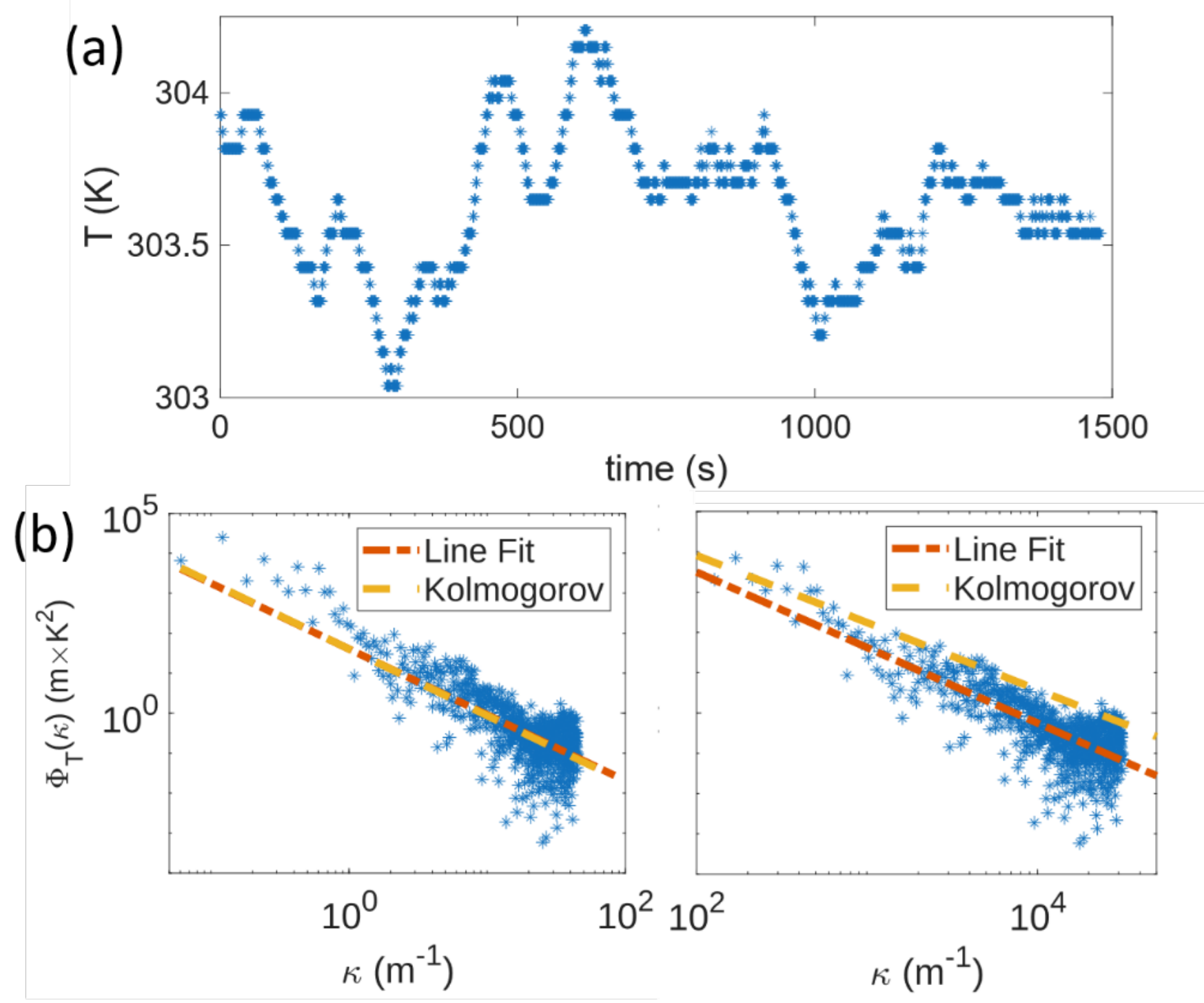

**Figure S2. (a)** $T(t)$ measurement. Each point is one sample. **(b)** $\Phi_T(\kappa)$ measurements overlaid with line fit and a line with slope $-5/3$ (1D Kolmogorov). *Left plot:* uses our measured upper limit on mean air speed ($\overline{v_y} = 0.07$ km/hour). The line fit has slope = $-1.65$ and mean RMS error = 1.39 $\text{m}\cdot\text{K}^2$. The line and Kolmogorov fits lie on top of each other. *Right:* Uses measured lower limit on mean air speed ($\overline{v_y}{\sim}10^{-4}$ km/hour). The line fit has slope = $-1.87$ and mean RMS error = $1.45\ \text{m}\cdot\text{K}^2$ .

To determine $\alpha$, we used a hot-wire anemometer to simultaneously measure 1D vertical air speeds (normal to the heater tape surface) and temperatures 2 cm above the heater tape, from which we computed the temperature power spectral density $\Phi_T(\kappa)$. First, we measured temperatures over time $T(t)$ for 10 minutes with measurement resolution of 1 s, as plotted in Fig. S2(a). Given mean vertical air speed $\overline{v_y}$ at measurement height off of the tape $y_{meas}$ and Taylor's Frozen Turbulence Hypothesis assuming static turbulent eddies moving vertically at $\overline{v_y}$ [4], this can be converted into $T(y) = T(t + y_{meas}/\overline{v_y})$ and used to compute $\Phi_T(\kappa)$. Then $\log(\Phi_T(\kappa))$ is fit to a line to determine the exponent $\alpha$ of the turbulence spectrum. The linear fits to the derived $\log(\Phi_T(\kappa))$ are shown in Fig. S2(b). Since the measured spectrum is that of 1D turbulence and not 3D, the Kolmogorov exponent is $\alpha = 5/3$ and not $\alpha = 11/3$, which is what is plotted for comparison in Fig. S2(b). Additionally, due to the extremely low vertical air speed ($v < 0.1$ km/h), we provided a line fit for the estimated lower limit and upper limits on mean speed in Fig. S2(b). The slightly higher $\alpha$, in this case $\alpha \sim 11.7/3$ for the lower limit $\overline{v_y}$, is consistent with weaker scintillation, as demonstrated in Section 3 of the main text. Despite this, all conclusions in the main text still hold, as they either depend solely on the standard definition of $C_n^2 = \langle (n(\mathbf{r} + \Delta\mathbf{r}) - n(\boldsymbol{r}) \rangle^2 / |\Delta\mathbf{r}|^{2/3}$ and not the details of the turbulence spectrum or account for this difference in turbulence spectrum.

## 4. Full calculation of turbulent scattering and waveguide efficiency

In Section 4 of the main paper, we reported calculated results on scattering of the guided beam out of an air waveguide by turbulence. This section describes that calculation, using results from [4].

The interaction between turbulent air and an optical beam can be described in terms of a series of statistical moments, the second-order moment being the mutual coherence function (MCF) between two transverse observation points $\mathbf{r_1}$ and $\mathbf{r_2}$, $\Gamma_2(\mathbf{r_1}, \mathbf{r_2}) = \langle E(\mathbf{r_1}) E^*(\mathbf{r_2}) \rangle = \iint E(\mathbf{r_1}) E^*(\mathbf{r_2}) d\mathbf{r_1} d\mathbf{r_2}$, where $E(\mathbf{r})$ is the scalar complex laser field. The MCF of an initially collimated Gaussian beam of spot size $w_0$ propagating a distance $L$ through Kolmogorov turbulence is [4]

$$\Gamma_2(\rho, L) = \Gamma_2(0, L) \exp\left[\frac{-\rho^2/w_0^2}{2\left(1 + \frac{4L^2}{(kw_0^2)^2}\right)} - \frac{3\sigma_R^2}{8}\left(\frac{1.22k\rho^2}{L}\right)^{5/6}\right], \tag{S5}$$

where $\rho = |\mathbf{r_1} - \mathbf{r_2}|$ is the separation between observation points. In the long-distance limit ($L \gg kw_0^2$), the second term in the exponential dominates. In our experiment with an f/750 focus ($w_0 \sim 0.25$ mm focal waist), this approximation holds well for the 1.8 m length of the heater tape.

We can use the MCF to calculate the average fraction of energy in the core region without the guide, $f_{ug,turb}$. The average fluence at $\mathbf{r}$ is equal to the MCF for $\mathbf{r} = \mathbf{r_1} = \mathbf{r_2}$ (or $\rho = 0$ in Eq. (S5)). The effective turbulence-induced beam waist is $w_{turb} = w_0(1 + 1.33\sigma_R^2(Lz_R/(z_R^2 + L^2))^{5/6})$ for $\sigma_R^2 = 1.23C_n^2 k^{7/6} L^{11/6}$ and $z_R = \pi w_0^2/\lambda$, such that the average intensity $\langle I(r, L) \rangle$ can be expressed as $\langle I(r, L) \rangle = w_0/w_{turb} e^{-2r^2/w_{turb}^2}$. Therefore, for a waveguide of core radius $a$,

$$f_{ug,turb} = \int_0^a re^{-2r^2/w_{turb}^2}\, dr \Big/ \int_0^\infty re^{-2r^2/w_{turb}^2}\, dr = 1 - e^{-2a^2/w_{turb}^2}. \tag{S6}$$

To calculate the average fraction of energy supported by the guide, $f_{g,turb}$, we take a 2D Fourier transform of the MCF with respect to $\rho$, which can be converted into a normalized distribution of transverse scattering angles $\theta$, which can be written as [4]

$$\widetilde{\Gamma_2}(\theta) = \left(\frac{8}{\pi\theta_{turb}^2}\right)^{1/2} e^{-2\theta^2/\theta_{turb}^2}. \tag{S7}$$

Here $\theta_{turb} = (3.099 C_n^2 k^{1/3} L)^{3/5}$ is the $1/e^2$ half-width of the scattering angular distribution for Kolmogorov turbulence. The distribution broadens as $\theta_{turb}$ increases from increasing turbulence and propagation length. Taking the waveguide as a cylindrically symmetric step-index structure, we consider the fraction of the beam energy confined by total internal reflection. This is the case for beam rays intersecting the cladding at angles below the critical acceptance angle of the waveguide, $\theta_{max} = \sin^{-1}(NA) = \sin^{-1}(\sqrt{n_{co}^2 - n_{cl}^2})$, where $n_{co}$ and $n_{cl}$ are the core and cladding indices of the waveguide. The fractional energy in the guide, $f_{g,turb}$, is the fraction of the unguided beam in the core, $f_{ug,turb}$, plus the fraction associated with scattered rays satisfying $\theta \leq \theta_{max}$. Therefore, the fraction of energy supported by the guide is

$$\begin{aligned} f_{g,turb} &= f_{ug,turb} + \left(1 - f_{ug,turb}\right)\sqrt{\frac{8}{\pi\theta_{turb}^2}}\int_0^{\theta_{max}} e^{-2\theta^2/\theta_{turb}^2}\,d\theta \\ &= f_{ug,turb} + \left(1 - f_{ug,turb}\right)\mathrm{erf}\left(\frac{\sqrt{2}\theta_{max}}{\theta_{turb}}\right) \end{aligned} \tag{S8}$$

For $\theta_{turb} \ll \theta_{max}$, turbulence is weak and $f_{g,turb} \to 1$. For $\theta_{turb} \gg \theta_{max}$, turbulence is strong and all light is scattered at large angles so that $f_{g,turb} \to f_{ug,turb}$. In these limiting cases, the fractional energy transmitted is independent of the guide.

Using the above results, we express the waveguide efficiency in terms of $f_{g,turb}$ and $f_{ug,turb}$ as $\eta_{turb} = \left[f_{g,turb} - f_{ug,turb}\right]/\left[1 - f_{ug,turb}\right]$, and assuming Kolmogorov turbulence we get

$$\eta_{turb}(L) = \mathrm{erf}\left[\frac{0.497\lambda^{1/5}\theta_{max}}{(C_n^2 L)^{3/5}}\right] \tag{S9}$$

For weak turbulence $(\theta_{turb} \ll \theta_{max})$, $\eta_{turb} \to 1$. For strong turbulence $(\theta_{turb} \gg \theta_{max})$, $\eta_{turb} \to 0$ as expected.

An estimate for the threshold level of turbulence where the beam is largely scattered out of the waveguide is found by requiring $\theta_{turb} < \theta_{max}$ so that the $1/e^2$ width of the scattering angular distribution remains inside the waveguide acceptance angle. This condition gives

$$\left(3.099 C_n^2 k^{1/3} L\right)^{3/5} < \sin^{-1}[(n_{co}^2 - n_{cl}^2)^{1/2}] \tag{S10}$$

or turbulence × length limit of

$$C_n^2 L < \frac{\lambda^{1/3}}{5.72}\left(\sin^{-1}[(n_{co}^2 - n_{cl}^2)^{1/2}]\right)^{5/3} \tag{S11}$$

Therefore, the threshold $C_n^2 L$ for strong scattering by turbulence depends only on the air waveguide numerical aperture and the guided light wavelength. These equations and results are discussed in Sec. 4 of the main text.